\documentclass{aa}  
\usepackage{booktabs}
\usepackage{tabularx}
\usepackage{makecell}
\usepackage{graphicx}
\usepackage{multirow}
\usepackage{array}
\usepackage{txfonts}
\usepackage{lipsum}
\usepackage{subcaption}  
\usepackage{longtable}
\usepackage{lscape}             
\usepackage{placeins}           
                                
\usepackage{hyperref} 
\hypersetup{
    colorlinks=true,
    linkcolor=blue, 
    filecolor=magenta,
    urlcolor=cyan,
    citecolor=blue, 
    pdftitle={TDR
},
    pdfauthor={S. Ronchini et al.},
}
\titlerunning{Targeted Detectability Range}
\authorrunning{Ronchini et al.}

\begin{document}
   \title{Gravitational-wave detectability range informed by external messengers}

   \author{S. Ronchini \inst{1,2}\fnmsep\thanks{ \email{samuele.ronchini@gssi.it}},
   A. Chopra\inst{1,2},
   T. Dal Canton\inst{3},
   B. Banerjee\inst{1,2},
   A. L. De Santis\inst{1,2},
   M. Branchesi\inst{1,2}
        }
   \institute{Gran Sasso Science Institute (GSSI), I-67100 L'Aquila, Italy
   \and
   INFN, Laboratori Nazionali del Gran Sasso, I-67100 Assergi, Italy
   \and
   Université Paris-Saclay, CNRS/IN2P3, IJCLab, 91405 Orsay, France}

   \date{}

  \abstract
   {A rapid estimate of gravitational-wave (GW) detectability associated with astronomical transients is crucial for optimizing multi-messenger follow-up strategies and for constraining the physical origin of the transient itself. We introduce the targeted detectability range (TDR), which is designed to evaluate with minimal computational effort the detectability of compact binary coalescences under the hypothesis of an association with an external messenger, such as an electromagnetic or neutrino signal. Unlike the standard GW range, which is based on averaged source parameters, the TDR incorporates prior information from observations of the external messenger, including sky localization, inclination constraints, and physically motivated bounds on the component masses. We report the TDR of all short- and long-duration gamma-ray bursts that were observed during the first three observing runs of Advanced LIGO and Advanced Virgo. The method was validated through a systematic comparison with the 90$\%$ exclusion distances provided by modeled targeted GW searches. In the absence of a coincident detection by all-sky all-time GW searches, the TDR provides a rapid and quantitative  constraint on a possible merger origin of the astrophysical source. Its low-latency implementation and public availability enable a timely prioritization of follow-up observations and optimized allocation of observational resources, with a direct impact on the physical interpretation of astronomical transients.}


   \maketitle
\section{Introduction}

Compact binary coalescences (CBCs) involving at least one neutron star (NS) can power the panchromatic electromagnetic (EM) emission from radio wavelengths to very high energies \citep{2020LRR....23....1M,2007PhR...442..166N}. During the first four observing runs of the LIGO–Virgo–KAGRA (LVK) collaboration, the interferometers achieved a sensitivity sufficient to probe binary neutron star (BNS) mergers up to 200-300 Mpc and neutron star-black hole (NSBH) mergers up to $\sim$1 Gpc \citep{lvr,2024A&A...686A.265C}. Within this volume, a wide variety of counterparts are also detectable in the electromagnetic domain, including prompt emission and afterglow of short gamma-ray bursts (GRBs) \citep{2014ARA&A..52...43B}, UV, optical, and infrared light associated with nuclear-decay-powered kilonovae \citep{1998ApJ...507L..59L,2020FrP.....8..355B}, soft and fast X-ray transients associated with emission produced outside the beaming cone of the GRB jet \citep{2017MNRAS.471.1652L,2020A&A...641A..61A}, high-energy radiation emitted by a shock breaking out from the post-merger ejecta \citep{2018MNRAS.475.2971B,2018MNRAS.473..576G,2019ApJ...871..200F,2025PhRvD.111f3031G}, and magnetar spindown-powered emission \citep{2013MNRAS.430.1061R,Chen_2026}. Merger-driven GRBs can also be potential sources of cosmic rays and high-energy neutrinos \citep{Guo2025}.
In this context, it is a critical task to assess whether a given non-GW trigger might be associated with a GW-detectable merger in low-latency (within about one hour from the trigger). Hereafter, whenever we refer to EM counterparts, the same considerations also apply to neutrino counterparts.

Within the LVK collaboration, 
the search for GWs targeted on external triggers, such as GRBs and high-energy neutrinos, employs the pipelines \texttt{PyGRB} \citep{2011PhRvD..83h4002H,2014PhRvD..90l2004W} and \texttt{X-Pipeline} \citep{2010NJPh...12e3034S,2012PhRvD..86b2003W}. \texttt{PyGRB} is a coherent matched filtering pipeline optimized for the search of CBC signals, while \texttt{X-Pipeline} is suitable for a search of unmodeled generic GW transients. These methods are computationally expensive and typically require a few hours of latency. A recent development of a targeted search version of the \texttt{GstLAL} pipeline shortened the computational time by a factor $\sim$50 compared to the time required by the all-sky all-time version of the search \citep{2026arXiv260622266T}.
Space- and ground-based follow-up resources, for example, deep optical spectroscopy, soft X-ray follow-up, radio monitoring, or very high-energy $\gamma$-ray observations, are limited and must be prioritized on timescales ranging from minutes to days from the EM trigger. The efficient allocation of these resources is becoming increasingly critical with the advent of wide field-of-view transient surveys, including missions such as the Einstein Probe \citep{2022hxga.book...86Y} in the X-ray band and the Vera Rubin Observatory \citep{2019ApJ...873..111I}. The latter is expected to operate as a powerful transient-discovery machine, substantially increasing the rate at which new extragalactic optical transients are identified \citep{2025ApJ...994L..24F}, some of which might be associated with CBCs \citep{2024arXiv241104793A}. In the absence of a GW candidate publicly released by LVK on GraceDB\footnote{\url{https://gracedb.ligo.org}}, which is temporally and spatially coincident with an EM trigger, the ability to rapidly assess whether a merger origin is compatible with the reach of operating GW interferometers is essential for prioritizing follow-up observations and interpreting the nature of the transient.

Currently, the LVK collaboration reports the status of each operating interferometer (IFO) in the form of a BNS range\footnote{\url{https://gwosc.org/detector_status/}} in real-time, computed for a $1.4-1.4$ $M_{\odot}$ equal-mass system and averaging over sky location and inclination of the orbital plane\footnote{\url{https://git.ligo.org/computing/services/gwistat/-/wikis/home/Range-Calculation}}. Notably, the maximum reachable distance, known as detection horizon, is achievable for the most favorable configuration of the binary parameters and is equal to $\sim $2.26 times the range \citep{FindChirp}.
While the range and the horizon provide a useful average characterization of the detector performance, the specific information given by the EM trigger is not considered. 

To overcome this limitation, we introduce the targeted detectability range (TDR), a real-time low-latency tool designed to evaluate the probability of detecting a hypothetical GW signal associated with an astrophysical transient under the assumption that it is produced by a CBC. In Sec. \ref{method} we describe the method we used to derive the TDR, exploiting the sensitivity of all the interferometers operating at the time of the EM trigger and incorporating prior information derived from the EM observation. In Sec. \ref{grb} we show the systematic application of this tool to all the GRBs detected during the first three observing runs of LVK and compare this with the  search results published offline. In Sec. \ref{cases} we show the application of TDR to possible science cases.

\section{Method}
\label{method}

We compute the detectability range starting from the sensitivity curve of each of the interferometers that operate in observing mode at the time of the EM trigger by exploiting the prior information we have from the astronomical source, including sky location, inclination angle (Sec. \ref{em_prior}) and component masses (Sec. \ref{sec:bns} and \ref{sec:nsbh_spin}) of the CBC. The TDR tool is publicly available\footnote{\url{https://github.com/samueleronchini/gw_tdr}}, with a user-friendly interface accessible via browser\footnote{\url{https://samueleronchini.github.io/gw_tdr/}}.

\subsection{Incorporating EM priors}
\label{em_prior}

The GW searches targeted on external triggers make use of restricted priors on the source parameters in order to increase their sensitivity. We followed a similar approach in defining the appropriate EM priors for the TDR calculation, according to the observational evidence coming from the external trigger. The first information is about the sky localization, whose uncertainty depends on the EM instrument. Typical sky localization areas range from arcseconds to arcminutes, for instance, in the case of optical or X-ray transients, to tens to hundreds of degrees for $\gamma-$ray sources localized by detectors such as Fermi-GBM. As explained below, the TDR is informed primarily by the sky localization map of the EM trigger.

Another relevant parameter that can be constrained a priori from the EM observations is the inclination angle between the CBC total angular momentum and the line of sight. If the EM transient is a GRB, it provides direct evidence of collimated radiation from a relativistic jet \citep{KUMAR20151}. The most reliable estimate of the aperture angle of short GRB jets is inferred from the identification of a temporal feature in the afterglow light curve, known as the jet break \citep{2002ApJ...571..779P}. Recent studies reported typical values of $\sim$10 deg, but higher values can be allowed and are just not measured due to detection biases \citep{2015ApJ...815..102F}. Therefore, when the EM candidate is associated with a relativistic jet, it is appropriate to conservatively assume a binary inclination in the range $\iota \in$ [0,45] deg. In the face-on configuration, the loudness of the GW signal is maximized because the S/N scales as $
\sqrt{\left(1+\cos ^2 \iota\right)^2+4\cos ^2 \iota}
$ (Eq. 3.31 of \citealt{1993PhRvD..47.2198F}). If the external EM event has the characteristics of a kilonova, we lack strong priors on the inclination of the binary because the angular distribution of outflows in an NS merger is expected to be nearly isotropic \citep{2020ApJ...889..171K}. In the case of soft, fast X-ray transients, the appropriate assumptions on the inclination angle depend on whether the source is more compatible with a relativistic jet origin, viewed either on- or mildly off-axis \citep{2020A&A...641A..61A,2026A&A...708A.190I}, or rather with more isotropic components, such as cocoon shock break-out or magnetar spin-down powered winds \citep{Chen_2026}. The multiwavelength observations following the detection of fast X-ray transients and the presence or absence of a simultaneous detection of an emission in the MeV band can help in favoring or disfavoring some of these scenarios and make a more tailored choice of the inclination angle \citep{2019Natur.568..198X,2024A&A...683A.243Q,2024A&A...690A.101W}. 

We worked under the assumption that the CBC is able to power an EM transient, and we therefore restricted the parameter space of the component masses by requiring  that a nonzero amount of baryon mass remains outside the merger remnant. We refer to this quantity as the remnant mass, to not be confused with the mass of the compact object formed after the merger. The remnant mass can be either bound, such as a hyper-accreting disk around the central remnant, or unbound. The unbound mass is a combination of several components, including dynamical ejecta (tidal tails in the equatorial plane and shock-driven ejecta at higher latitudes), neutrino-driven winds from the remnant and accretion disk, viscous (secular) disk outflows powered by angular momentum transport and nuclear recombination, and magnetically driven winds and polar outflows associated with the remnant and jet-launching region \citep{doi:10.1142/S021827181842004X,2018ApJ...869..130R,2020ARNPS..70...95R,2020FrP.....8..355B,2020LRR....23....1M}. Additional ejecta can be produced as a particle wind when the merger remnant is a fast-rotating magnetar \citep{2013ApJ...776L..40Y,2021ApJ...912...14Y}. A nonzero ejected mass can also allow the production of other messengers in addition to GWs and photons, such as high-energy neutrinos and ultrahigh-energy cosmic rays (see Sect. \ref{sec:Neutrino}).

\subsubsection{BNS parameters required to produce EM emission}
\label{sec:bns}
In the case of a BNS merger,  the ability to power an EM counterpart primarily depends on the component masses and the nature of the merger remnant. When the total mass is sufficiently high, the merger undergoes a prompt collapse to a BH.  Conversely, lower-mass systems can produce a stable or short-lived metastable NS, surrounded by a massive accretion disk (up to $\sim$$10^{-1}M_\odot$) and substantial ejecta, able to power GRB and kilonova emission. The mass of the accretion disk and dynamical ejecta depends on the mass ratio, which is more massive for more asymmetric systems \citep{2019ARNPS..69...41S,2020MNRAS.497.1488B}. High accretion disk masses have been proposed to interpret the long-duration merger-driven GRBs \citep{2023ApJ...958L..33G}.

When the jet is launched by the Blandford-Znajek \citep{1977MNRAS.179..433B} or neutrino-antineutrino annihilation \citep{2011MNRAS.410.2302Z} processes, then an accretion disk is a necessary ingredient. However, even in the absence of accretion, a GRB jet might be powered by the spin-down energy of a millisecond magnetar created after the merger \citep{2011MNRAS.413.2031M,2012MNRAS.419.1537B,2026arXiv260611299K}, although this scenario is still debated and it is still unclear whether a proto-magnetar can produce such a jet \citep{2020MNRAS.495L..66C,2023ApJ...947L..15M}. 
The spin-down of a newborn magnetar can create a wind \citep{2013ApJ...776L..40Y,2021ApJ...912...14Y}, which, by shocking the circum-burst medium, can power a fast X-ray transient \citep{2019ApJ...886..129S,2024A&A...683A.243Q} or appear as an extended emission (plateau) in the $\gamma-$ray (X-ray) light curve of GRBs \citep{2013MNRAS.431.1745G,2013MNRAS.430.1061R,2018ApJ...869..155S,2023A&A...675A.117R,2025NSRev..12E.401S}.

The kilonova emission, instead, is powered by multiple components of unbound ejecta, including the dynamical ejecta, the neutrino-driven wind ejecta, and the secular ejecta, which originate from matter unbound from the accretion disk on longer timescales \citep{2016MNRAS.463.2323W,2017PhRvL.119w1102S}. While the brightness of the kilonova can be limited by small dynamical ejecta, due to prompt BH collapse and mass ratios $\sim$1 \citep{2019ARNPS..69...41S}, a magnetar remnant can significantly enhance the kilonova brightness \citep{2014MNRAS.439.3916M,2022MNRAS.516.4949S,2025ApJ...978...52A}.

In summary, this plethora of scenarios shows that a wide range of NS masses and mass ratios can potentially power an EM emission. Motivated by this, we did not apply any stringent cut on the NS component masses, allowing the TDR user to choose the mass configuration best suited to the scenario of interest. In order to explore the widest range of chirp mass, the adopted combinations of masses are $[M_1, M_2] = [1,1]M_{\odot}$, $[1.4,1.4]M_{\odot}$, and $[2.0,2.0]M_{\odot}$.

\subsubsection{NSBH parameters  required to produce EM emission}
\label{sec:nsbh_spin}

\begin{figure}
    \centering
    \includegraphics[width=1\linewidth]{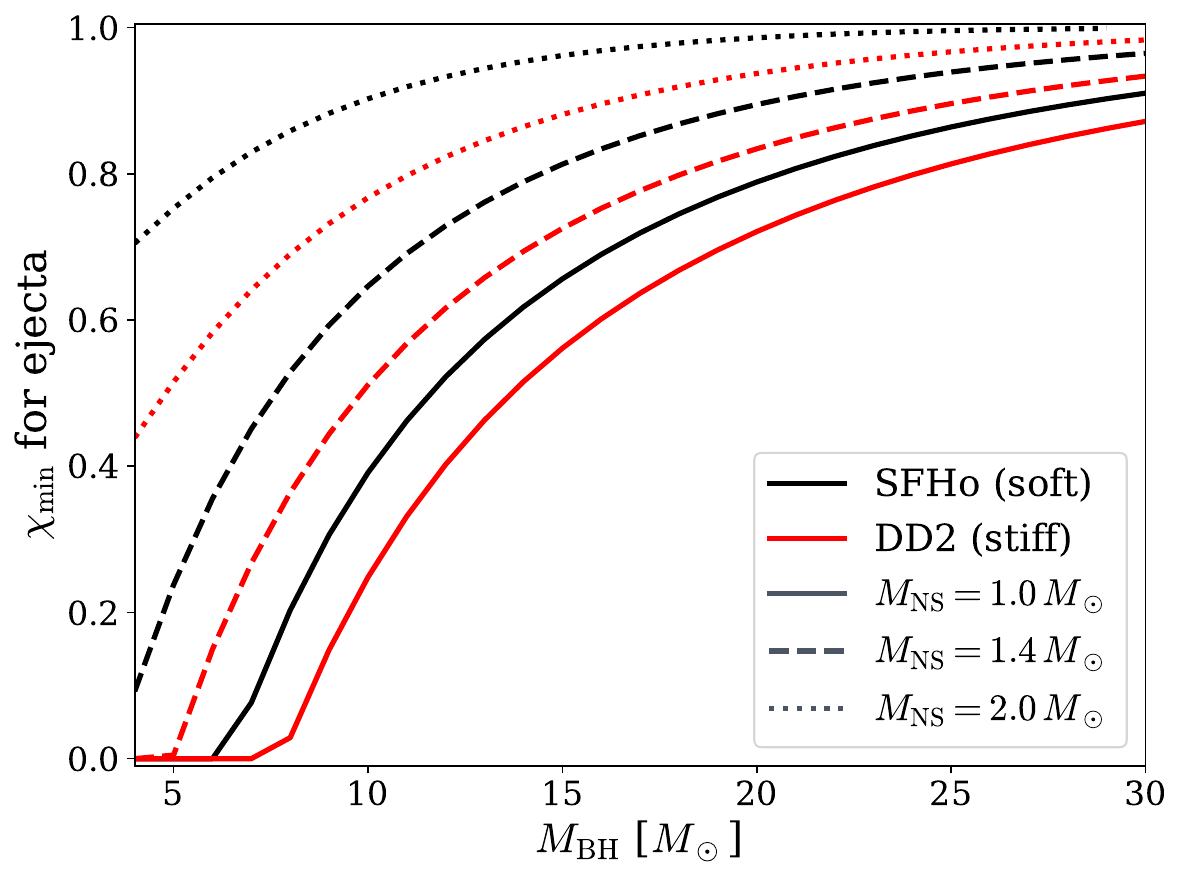}
    \caption{Minimum BH spin $\chi_{\rm min}$ required to have $M_{\mathrm{REM}}>0$ as a function of BH mass for $M_{\mathrm{NS}}= 1.0, 1.4, 2.0 \,M_{\odot}$, evaluated for the SFHo and DD2 equations of state.}
    \label{fig:foucart}
\end{figure}

The probability of a nonzero remnant mass in an NSBH merger is primarily a function of the mass ratio, NS equation of state (EOS), and BH spin \citep{2011ApJ...727...95P,2012PhRvD..86l4007F}. Hereafter, we refer to the fitting formulae reported in \citet{Foucart2018} for the estimation of the remnant mass. We considered three representative values of NS masses, corresponding to 1 $M_{\odot}$, 1.4 $M_{\odot}$, and 2 $M_{\odot}$. This range is compatible with the Galactic population of NS and also with the maximum NS mass allowed by the EOS \citep{2016ARA&A..54..401O}. As reference for the EOS, we chose the soft SFHo EOS \citep{HSNP_2010, SHF_2013} and the stiff DD2 EOS \citep{HSNP_2010, TRKBP_2010}. 
These EOSs satisfy current constraints from GW170817 and the associated kilonova (see, e.g., \citealt{2018ApJ...852L..29R}) and encompass the current range of uncertainties in NS compactness.
 
For each fixed NS mass and EOS, a direct correspondence exists between the BH mass and the minimum BH spin required to have a remnant mass $M_{\mathrm{REM}} > 0$. 
A higher spin results in a higher BH mass. Therefore, in order to estimate the TDR for the widest range of chirp mass, we explored three spin ($\chi_{BH}$) regimes: a low-mass nonrotating BH with $\chi_{BH}=0$ for the lowest NS mass of 1 $M_{\odot}$, a BH with an intermediate value of $\chi_{BH} \sim 0.5$ for the NS mass of 1.4 $M_{\odot}$, and a larger BH with an extreme value of $\chi_{BH} > 0.9$ for the NS mass of 2 $M_{\odot}$. Fig.~\ref {fig:foucart} shows the minimum BH spin required to have $M_{\mathrm{REM}}>0$ as a function of the BH mass and fixing the NS mass to our representative values of 1.0, 1.4, and 2.0 $M_{\odot}$ for the SFHo and DD2 EOSs \citep{Hinderer2008, Damour2009}. Following these prescriptions, we arrive at the choice of the following mass combinations: 
\begin{enumerate}

    \item $[M_{\mathrm{NS}}, M_{\mathrm{BH}}] = [1,5]M_{\odot}$
    \item $[M_{\mathrm{NS}}, M_{\mathrm{BH}}] = [1.4,10]M_{\odot}$
    \item $[M_{\mathrm{NS}}, M_{\mathrm{BH}}] = [2,20]M_{\odot}$.
\end{enumerate}
The minimum BH spins required to leave baryonic mass outside the merger remnant for our choice of component masses are listed in Tab.\,\ref{tab:spin_thresholds}. In summary, for the $[M_{\mathrm{NS}}, M_{\mathrm{BH}}] = [1,5]M_{\odot}$ combination, any BH spin gives $M_{\mathrm{REM}}>0$, for $[M_{\mathrm{NS}}, M_{\mathrm{BH}}] = [1.4,10]M_{\odot}$ an intermediate spin in the range $\chi_{BH}\sim 0.5-0.6$ is required, and for the $[M_{\mathrm{NS}}, M_{\mathrm{BH}}] =[2,20]M_{\odot}$ combination, a high spin value of ($\chi_{\mathrm{BH}}\sim 0.94-0.99$) is required. For completeness, we explicitly list in Tab.\,\ref{tab:spin_thresholds} the NS tidal deformability $\Lambda$ in each scenario as this is an EOS-dependent parameter used within our considered waveforms. We computed $\Lambda$ by numerically inverting the universal relation between compactness and tidal deformability derived by \citet{Godzieba2021}.  
According to GW observations \citep{2026ApJ..1004L..21A,2025arXiv250818083T} and theoretical models of CBC populations \citep{2022ApJ...928..163H}, there is still no full consensus about the expected value of the BH spin in NSBH systems. When the binary comes from an isolated channel, the magnitude of the BH spin depends on whether the BH forms before or after the companion \citep{2020A&A...635A..97B,2019ApJ...870L..18Q}. Moreover, there is observational evidence of highly spinning BHs in X-ray binaries, which might eventually evolve into merging NSBH systems \citep{2019IAUS..346..426Q}. For a binary inside a stellar cluster, a low BH spin is expected when the BH forms from an isolated star, unless there are repeated episodes of mass accretion throughout the evolution of the BH \citep{2019ApJ...881L...1F}. Therefore, the presence of NSBH systems with BH spins as high as those assumed for the mass combination $[M_{\mathrm{NS}}, M_{\mathrm{BH}}] =[2,20]M_{\odot}$ is still a viable scenario, even though it is considered more exotic than the less massive combinations.

\subsubsection{High-energy neutrino triggers}\label{sec:Neutrino}

The TDR framework is directly applicable to external triggers from high-energy  neutrino observatories. 
High-energy neutrinos are expected to be produced in the relativistic jet 
launched by BNS and NSBH mergers \citep{Kimura2017, Guo2025}.
Previous searches for joint GW and neutrino emission in all LVK observing runs, using IceCube \citep{Aartsen2020, Abbasi2023a,  Abbasi2023b,2024PhRvD.110f3004M,2025ApJ...987..218M}, ANTARES \citep{Albert2017, Albert2020, Albert2023}, and KM3NeT  \citep{Aiello2024}, yielded no significant detections to date. The TDR provides  the complementary neutrino-triggered perspective: given a real-time alert from  a facility such as KM3NeT/ARCA \citep{AdrianMartinez2016}, it rapidly assesses whether an associated GW signal would have been detectable by the operating interferometers. 
The choice of inclination prior depends on the inferred emission scenario: if the neutrino is consistent with prompt emission from a relativistic jet, the GRB case applies ($\iota \in [0, 45]$ deg); 
otherwise, for instance, in 
the case of a choked jet where the neutrino escapes but no GRB is observed, an isotropic prior ($\iota \in [0, 90]$ deg) should be adopted. 
In the absence of an EM counterpart, it is advisable to run the TDR for all available mass combinations and both inclination priors. The sub-degree angular resolution of track-like neutrino events at energies above a few tens of TeV makes the sky-localization contribution to the antenna-factor uncertainty negligible. The main additional complication with respect to GRB triggers is the uncertain delay between merger time and neutrino emission, which can range from seconds to days. The neutrino emission can indeed occur either during the prompt phase or simultaneously with the afterglow emission. This can be handled by evaluating the TDR sequentially over multiple time segments around the neutrino trigger time, as described in Sec.~\ref{conc}.

\subsection{Computation of optimal S/N and $D_{90}$}
\label{suppl}

In order to simulate the signal of the CBC inspiral, we used the PyCBC \citep{2016CQGra..33u5004U} module \texttt{pycbc$\_$create$\_$injections}. We used the \texttt{TaylorF2} \citep{PhysRevD.75.124018} waveform for BNS signals and \texttt{IMRPhenomNSBH} \citep{PhysRevD.100.044003} for NSBH. We set the NS spin and BH tidal deformability to zero, while the tidal deformability $\Lambda_2$ of the NS was taken from Tab. \ref{tab:spin_thresholds} for each NS mass and EOS\footnote{Since the TDR for NSBH depends slightly on the EOS, we report $D_{90}^{\mathrm{TDR}}$ as the average obtained using the SFHo and DD2 EOSs.}. The black hole spin was fixed to the corresponding minimum value necessary for a remnant mass $M_{\mathrm{REM}} > 0$, listed in Tab.~\ref{tab:spin_thresholds}. Polarization angle and coalescence phase angle were distributed uniformly in the range $[0,2\pi]$. 
Since the NS and BH mass distributions are still poorly constrained, we adopted fixed component masses (see Sec.~\ref{sec:bns} and \ref{sec:nsbh_spin}) that spanned the relevant mass parameter space for BNS and NSBH. This means that the TDR tool by default produced results for a total of six (three plus three) mass combinations.
As motivated in Sec.~\ref{em_prior}, we computed the TDR under two assumptions: 1) an isotropic distribution in the range $\iota \in [0,\pi/2]$, and 2) an isotropic distribution in the range $\iota \in [0,\pi/4]$. 
The sky localization of the EM transient was taken into account as follows. When the source was localized with a precision of arcminutes to arcseconds, the RA and Dec of the injections were kept fixed. When the source is not well localized, such as GRBs detected by Fermi-GBM, or was obtained with InterPlanetary Network (IPN) triangulation \citep{2013ApJS..207...38P,2020ApJ...895...40G,2022ApJS..259...34S,2025ApJS..278...60X}, we distributed the RA and Dec according to the probability distribution included in the sky localization map of the EM source. 

The optimal S/N of each injected source was computed using the \texttt{pycbc$\_$optimal$\_$snr} module from PyCBC. The optimal S/N is defined as
\begin{equation}
\rho_{\mathrm{opt}}=\left [4 \int_{f_\mathrm{L} }^{f_\mathrm{H} } \frac{|\tilde{h}(f)|^2}{S_n(f)} d f \right]^{1/2},
\end{equation}
where $\tilde{h}(f)$ is the Fourier transform of the GW signal, and $S_n(f)$ is the power spectral density (PSD) of the detector noise. We adopted $f_\mathrm{L} = 30$ Hz and $f_\mathrm{H} = 1024$ Hz. The choice was made to be consistent with the values adopted in the GRB targeted searches performed by LVK (sec. 3.1 in \citealt{2021ApJ...915...86A}). Outside this frequency range, the contribution to the S/N is negligible because of the current sensitivity of GW detectors. If the TDR were used on future GW detectors, this range should be extended accordingly.
To reduce the computational cost of S/N estimation, the only parameters randomized during injection generation were inclination angle, polarization angle, and phase at coalescence. This yielded a probability distribution of the S/N at a fixed reference luminosity distance and sky position. We call $\rho_{\mathrm{ref},n}$ the collection of optimal S/N calculated at a reference distance $D_{L, \rm ref}$ and sky position $(\text{RA}_{\rm ref}, \text{Dec}_{\rm ref})$ for an optimally oriented binary ($\iota=0$) and for a reference polarization angle $\psi=0$. It suffices to know the S/N distribution as a function of distance and sky position to rescale each S/N of the reference  distribution using the relation
\begin{equation}
\rho_n(\text{RA}, \text{Dec}, D_L,\iota, \psi)=\rho_{\mathrm{ref},n} \frac{D_{\text {eff }, \mathrm{ref}}}{D_{\text {eff }}(\text{RA}, \text{Dec}, D_L,\iota, \psi)},
\end{equation}
where
\begin{equation}
D_{\mathrm{eff}}(\text{RA}, \text{Dec}, D_L,\iota, \psi)
= D_L \Bigg[
F_{+}^2\left(\frac{1+\cos^2 \iota}{2}\right)^2
+ F_{\times}^2 \cos^2 \iota
\Bigg]^{-1/2}
\end{equation}
and
\begin{equation}
D_{\mathrm{eff,ref}}=D_{\text {eff }}(\text{RA}_{\rm ref}, \text{Dec}_{\rm ref}, D_{L, \rm ref},\iota=0, \psi=0),
\end{equation}
where $D_L$ is the luminosity distance, and $F_{+}$,$F_{\times}$ are the plus and cross components of the antenna pattern of the interferometer \citep{FindChirp}. The dependence on $\psi$ is enclosed in $F_{+}$,$F_{\times}$. Finally, the optimal network S/N is 
\begin{equation}
\mathrm{\rho_{net}} = \sqrt{\sum_i \rho^2_i},
\end{equation}
where the sum runs over all the interferometers online. After we obtained the distribution of the network S/N, the EM informed detectability range $D_{\lambda}|_{\rho_{\mathrm{cut}}}$ was defined as the distance where
\begin{equation}
P(\mathrm{\rho_{net}}>\rho_{\mathrm{cut}}) = \lambda,
\end{equation}
namely where a fraction $\lambda$ of the injected sources are recovered with an optimal S/N$ > \rho_{\mathrm{cut}}$. 

The value $\rho_{\mathrm{cut}}$ we adopted for the S/N cut is purely representative and can be adjusted by the user according to the specific application. Since the optimal S/N is a monotonic function of distance, at distances larger than $D_{\lambda}|_{\rho_{\mathrm{cut}}}$, fewer than a fraction $\lambda$ of the sources are detected with S/N$ > \rho_{\mathrm{cut}}$. 
As defined in \cite{2017ApJ...841...89A,2019ApJ...886...75A,2021ApJ...915...86A,2022ApJ...928..186A}, GW targeted searches report the $90\%$ exclusion distance using a frequentist approach, defined as the distance at which 90$\%$ of injections are recovered with a ranking statistics larger than the maximum observed in the on-source time window. Analogously, we defined the TDR, $D_{90}$, choosing $\lambda=90\%$.
Since the source S/N and detection significance are strongly correlated, the TDR is also expected to correlate on a statistical basis with the exclusion distance derived by LVK targeted searches.

The status of each interferometer was checked using the \texttt{dq.query$\_$flag} from PyCBC, which identifies all the time segments during which the instrument was taking data. The single interferometer was considered to be online and was therefore included in the estimation of the network S/N when the data coverage lasted the entire time interval $[t_0-128s,t_0+128s]$, where $t_0$ is the EM trigger time. This specific window was defined as the minimum duration required to ensure a reliable estimate of each IFO noise level. Since this tool can run in low-latency, it is easily scalable and the estimation of the range can be repeated in multiple time intervals even when the EM-GW time delay is highly uncertain. The TDR tool is designed to work in real time and low latency, provided that the PSD of each IFO is promptly available. However, the LVK collaboration does not release the frames for a PSD estimation in real-time data, but just day-averaged estimates\footnote{e.g. \url{https://gwosc.org/detector_status/day/20250716/}}. This implies that the TDR can only be run offline by the user after the data release. Therefore, we show the use of the TDR using strain data files publicly available in the GWOSC database\footnote{\url{https://gwosc.org}}, even though the implementation for an online low-latency analysis is immediate. The name of the strain channel used for each IFO and each observing run is reported in Tab.~\ref{runs}. The noise PSD was computed from the strain time series using Welch’s method \citep{WelchPSDEstimation}, a widely adopted technique for PSD estimation in time-series analysis, as it reduces the variance of the estimate by averaging periodograms of overlapping segments. In this implementation, each segment has a length of 16 s, with 50$\%$ overlap between consecutive segments. This balances spectral resolution and statistical stability.

\subsection{Accounting for stationary Gaussian noise and waveform mismatch}
\label{meatch}

In the method described so far, we only used the concept of an optimal S/N, which approximates the average expected matched-filter S/N under the assumption of stationary Gaussian noise and a perfect match between the true GW signal and the template (model) waveform used by the search pipeline. The optimal S/N is uniquely given by the signal model and the noise PSD. However, the matched-filter S/N that is derived by search pipelines is the result of a comparison between a template and the real data, the latter being the sum of the astrophysical signal and noise. In terms of equations, when we decompose the data as the sum of GW strain plus noise, $s = h_\mathrm{true} + n$, then the optimal S/N can be written as
\begin{equation}
    \rho_{\mathrm{opt}} = \sqrt{(h_\mathrm{true} \mid h_\mathrm{true})},
\end{equation}
while the matched-filter S/N is
\begin{equation}
   \rho_{\mathrm{MF}}=\frac{\sqrt{(s \mid h_\mathrm{I})^2 + (s \mid h_\mathrm{Q})^2}}{\sqrt{(h_\mathrm{I} \mid h_\mathrm{I})}},
\end{equation}
where $h_\mathrm{I,Q}$ are two versions of the template waveform with a 90 deg phase difference between them, and the inner product $(\cdot|\cdot)$ is defined as
\begin{equation}
(a \mid b)=4 \Re \int_{f_\mathrm{L} }^{f_\mathrm{H} } \frac{\tilde{a}(f) \tilde{b}^*(f)}{S_n(f)} d f.
\end{equation}
When we consider all possible realizations of stationary Gaussian noise, $\rho_\mathrm{MF}^2$ is a random variable drawn from a non-central $\chi^2$ distribution with 2 degrees of freedom and non-centrality parameter $k \rho_\mathrm{opt}^2$,
where $k$ is a random variable modeling the mismatch between the true signal and template waveform, which we took to be uniformly distributed in the range $[0.9-1]$ \citep{2017arXiv170501845D,2026ApJ..1004L..21A}.
The above expressions are valid for a single detector; when we consider the network S/N (either optimal or matched-filter), the same expressions hold, but the $\chi^2$ distribution has $2 N_\mathrm{IFO}$ degrees of freedom instead, where $N_\mathrm{IFO}$ is the number of observing detectors. Hence, along with the $D_{90}$ distance defined above, adopting the cut $\rho_{\mathrm{opt}}>\rho_{\mathrm{cut}}$, we analogously derived the distribution of $D_{90}^{\mathrm{TDR}}$ by converting $\rho_{\mathrm{opt}} \rightarrow \rho_{\mathrm{MF}}$ using the $\chi^2$ distribution and again using the cut $\rho_{\mathrm{MF}}>\rho_{\mathrm{cut}}$.

\begin{figure}
    \centering
    \includegraphics[width=1\linewidth]{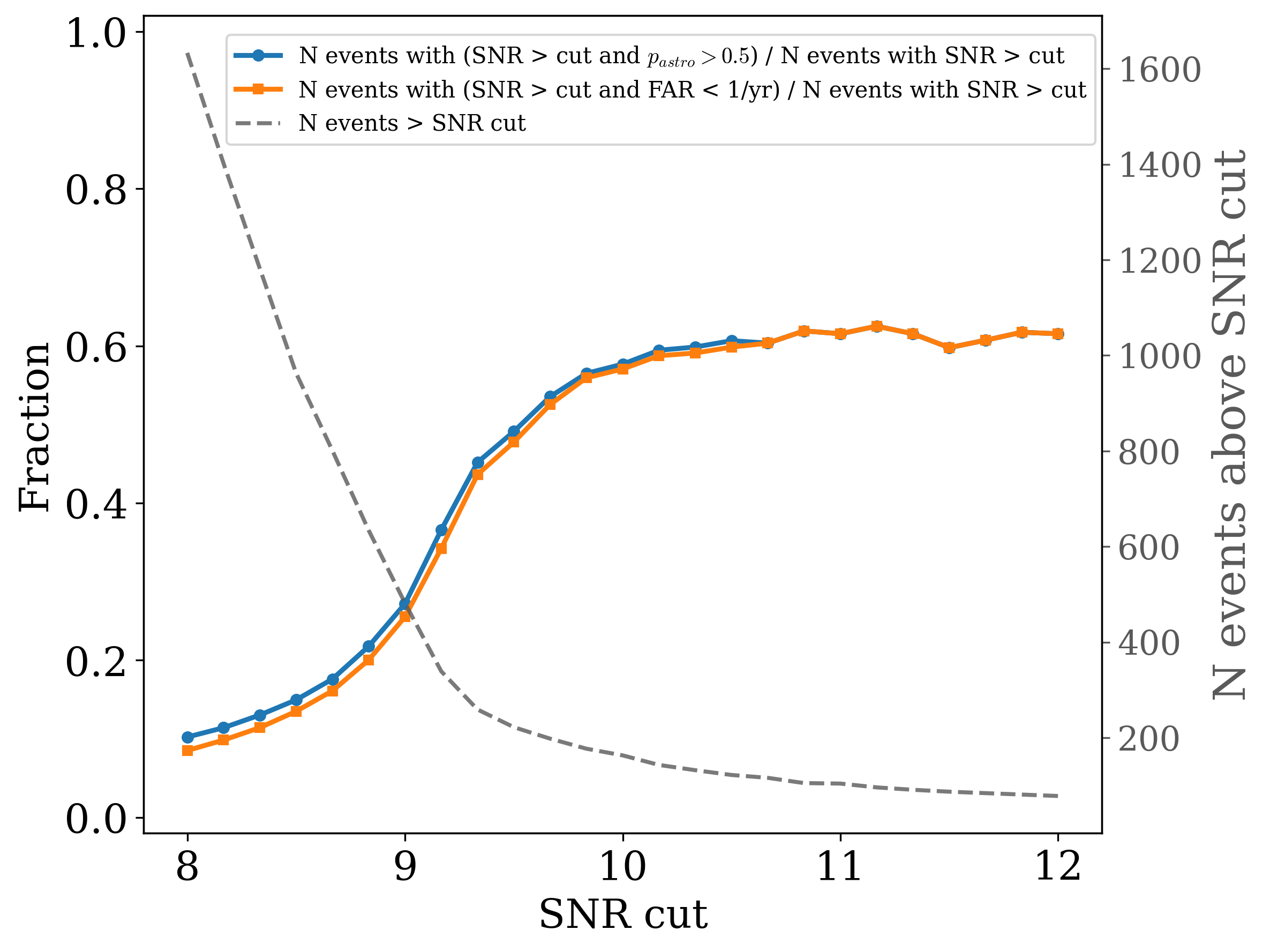}

    \caption{S/N cut purity considering all the GW candidates released by LVK. The candidates were taken from GWTC2.1, GWTC3, and GWTC4. The blue and orange lines indicate from all the events above the S/N cut the fraction of events with $p_{\mathrm{astro}}>0.5$ and FAR < 1 / yr, respectively (left y-axis). The dashed gray line indicates the number of all the events above the S/N cut (right y-axis).}
    \label{fig:purity}
\end{figure}

It is important to note that the $D_{90}^{\mathrm{TDR}}$ cannot be rigorously interpreted as an exclusion distance on its own because no actual search was performed here (i.e., we did not calculate the matched-filter S/N). However, $D_{90}^{\mathrm{TDR}}$ approximates an exclusion distance when no GW search pipeline finds any significant GW candidate in a given temporal window centered around the EM trigger. The choice of the S/N cut for the computation of $D_{90}^{\mathrm{TDR}}$ can be informed by the relation between the S/N and the significance of GW candidates found so far. The significance of candidates found by GW searches is quantified by the false-alarm rate (FAR) and the probability of being astrophysical by $p_{\mathrm{astro}}$. By searching the GWTC catalogs containing candidates found by offline analyses, it is possible to verify that a cut at S/N = 10 implies a fairly high purity of the candidate. This is shown in Fig.~\ref{fig:purity}, where we show that based on a selection of all the GW candidates found so far with an S/N $>$10, around 60$\%$ of them have an FAR$<$ 1/yr and a $p_{\mathrm{astro}}>0.5$. Equivalently, an astrophysical GW source found by search pipelines with a matched-filter S/N$>$10 has a moderately low probability to have an FAR that is high enough to be discarded as non-astrophysical, and hence, to be not reported as a viable candidate by low-latency searches or offline searches.

In conclusion, for a chosen value of S/N cut $\rho_{\mathrm{cut}}$, the resulting targeted detectability range, $D_{90}^{\mathrm{TDR}}$ is defined as follows:
given the detection of an external messenger, in the assumption of being driven by a CBC, there is a 90$\%$ probability at the distance $D_{90}^{\mathrm{TDR}}$ that the GW signal is found by a GW search pipeline with a measured S/N equal to or higher than $\rho_{\mathrm{cut}}$. The adopted value of $90\%$ is conservative, but purely representative, and was chosen to be consistent with the definition of the exclusion distance given by LVK targeted searches. However, since the TDR tool also provides the detection efficiency curve as a function of distance, the user can adopt a different detection efficiency cut according to the specific goal.

\section{GW detectability of GRBs detected during the first three LVK observing runs}
\label{grb}

\begin{figure}
    \centering
    \includegraphics[width=\linewidth]{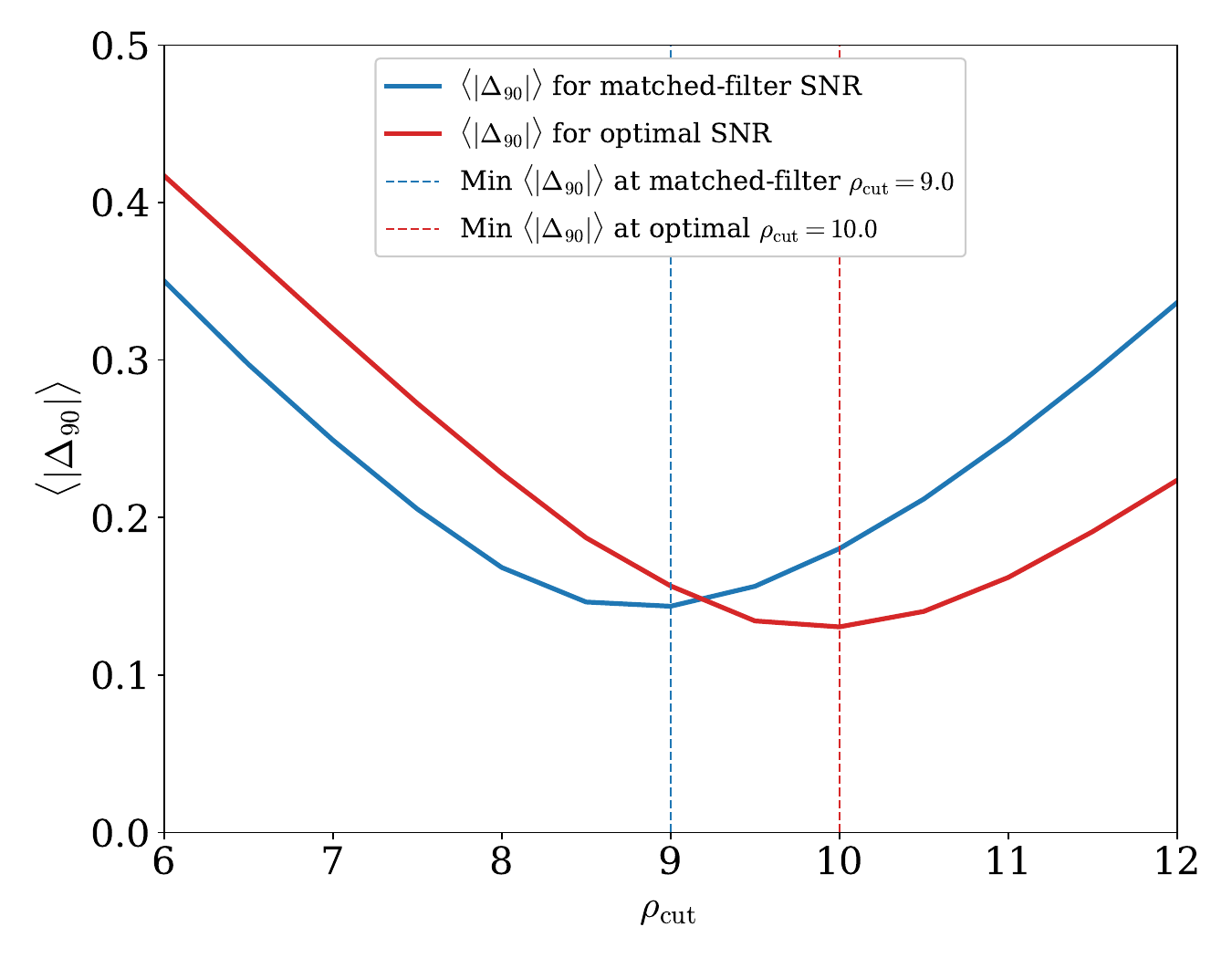}
    \caption{Average over all the analyzed GRBs of the relative difference between the TDR and the 90$\%$ exclusion distance obtained by \texttt{PyGRB} as a function of the S/N cut. The blue curve shows matched-filter S/N, and the red curve shows the optimal S/N.}
    \label{fig:minim}
\end{figure}

\begin{figure}
    \centering
    \includegraphics[width=1\linewidth]{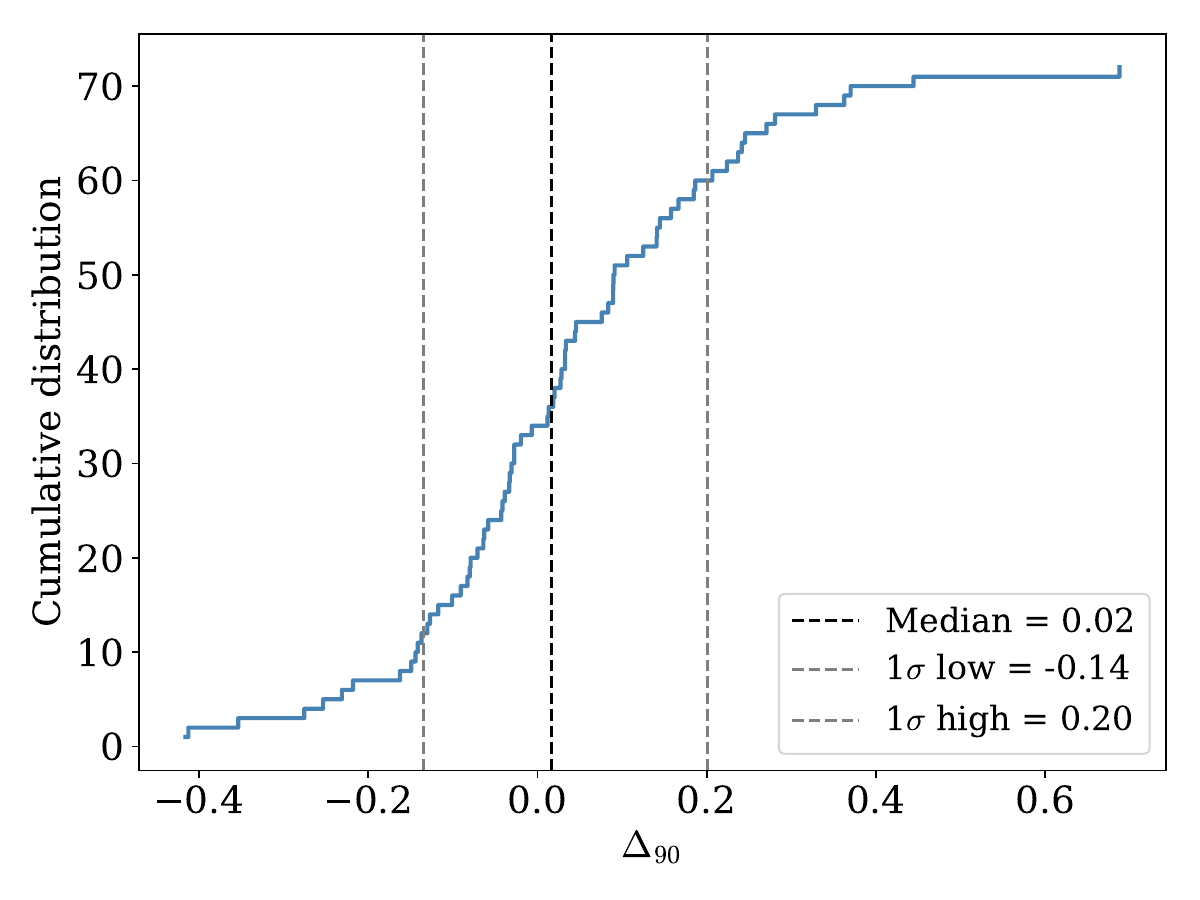}
    \caption{Cumulative distribution of the relative discrepancy between the TDR computed in this work and the $90\%$ exclusion distance found by \texttt{PyGRB} on all the GRBs analyzed between O1 and O3. The dashed lines indicate the 16th, 50th, and 84th percentiles of the distribution.}
    \label{fig:cumul}
\end{figure}
\begin{figure*}[t]
    \centering
    \includegraphics[width=.95\linewidth]{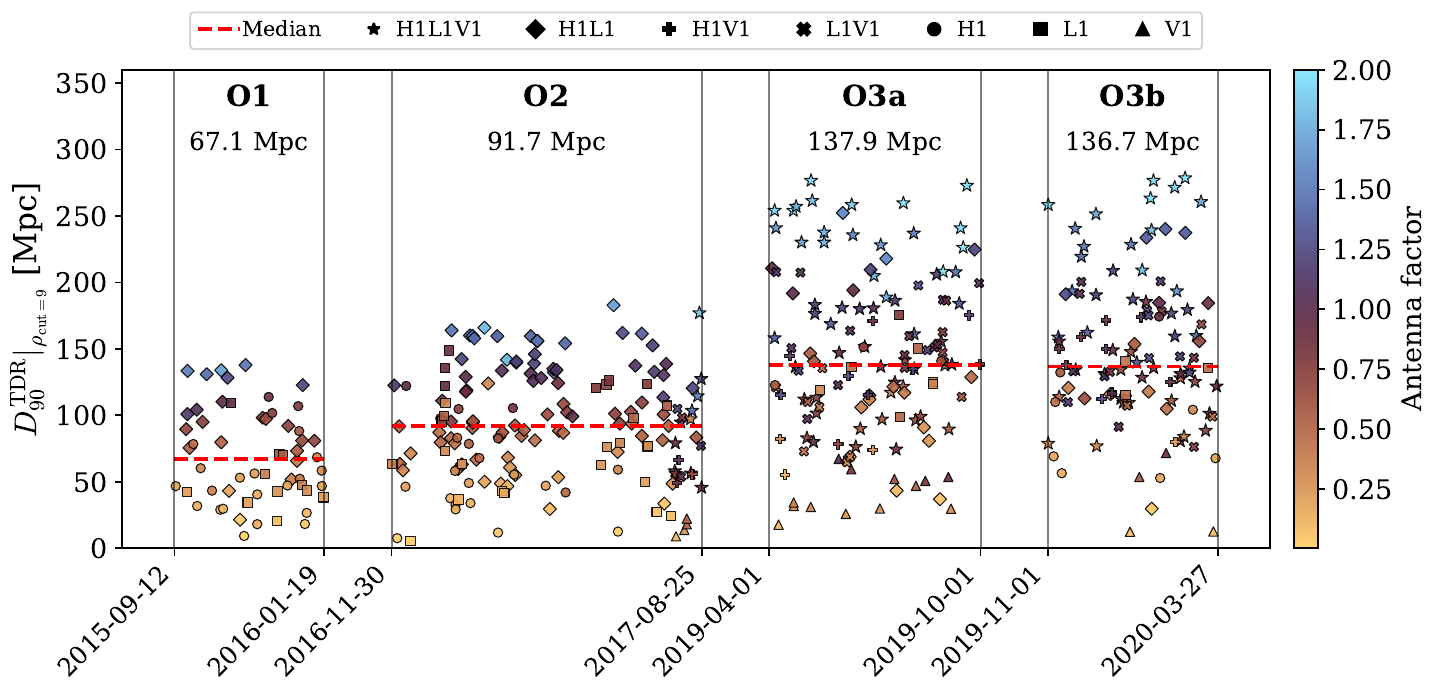}
    \caption{Targeted detectability range ($D_{90}^{\mathrm{TDR}}$) for all the GRBs reported by Fermi-GBM and Swift-BAT during the first 3 LVK observing runs. The range is computed for a BNS system with $m_1=m_2=1.4 M_{\odot}$, assuming an inclination of the binary isotropically distributed between 0 and 30 deg. Each marker indicates the IFOs taking data at the GRB trigger time, and the color is the network antenna factor at the GRB position. The median values of $D_{90}^{\mathrm{TDR}}$ are reported below each run and are indicated with a horizontal dashed red line.}
    \label{fig:time_evo}
\end{figure*}
\begin{figure}[t!]
    \centering
     \includegraphics[width=0.85
     \linewidth]{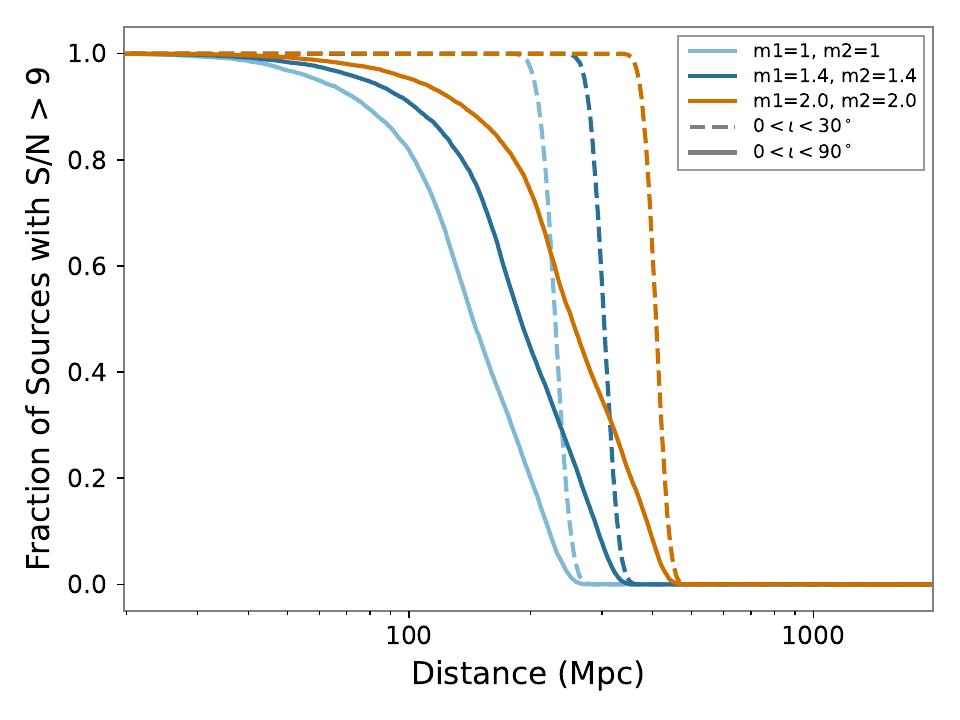}
     \includegraphics[width=0.85\linewidth]{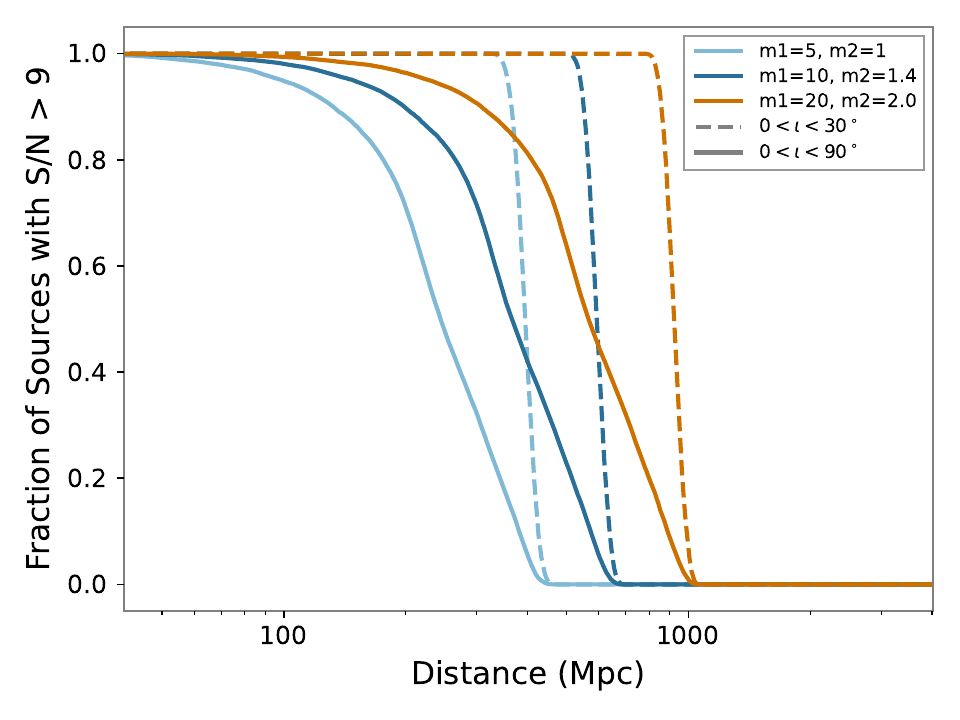}
    \includegraphics[width=0.95
    \linewidth]{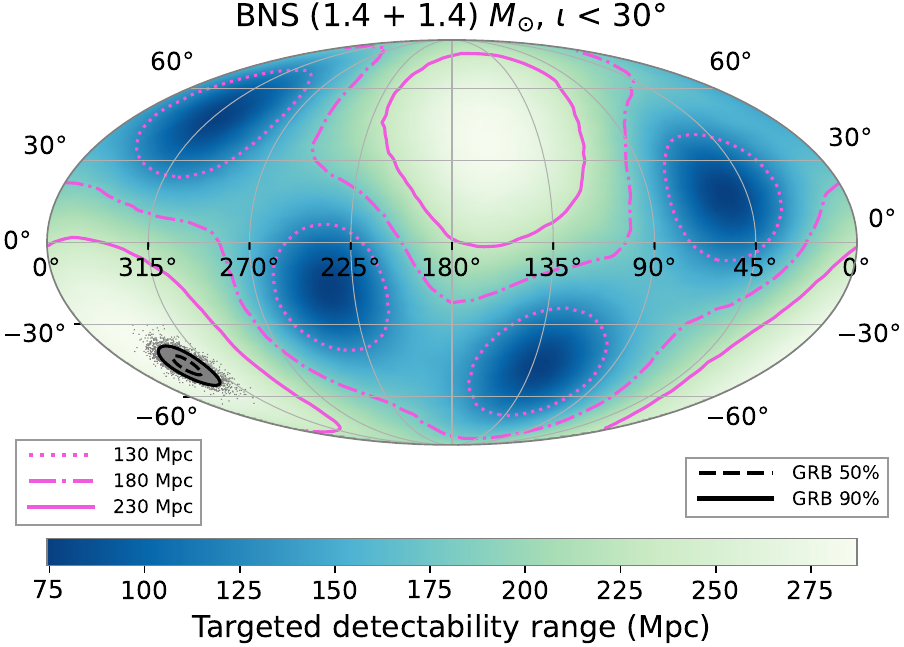}
    \caption{Top panel: Fraction of sources with matched-filter S/N > 9 as a function of luminosity distance for different equal-mass BNS systems, oriented nearly face-on (dashed lines) or randomly oriented (solid lines). Middle panel: Same as the top panel, but for NSBH, showing an average of the range for SFHo and DD2 EOSs. Bottom panel: Value of $D_{90}^{\mathrm{TDR}}$ as a function of the sky position. Equal-value levels (in purple) follow the directional dependence of the antenna pattern. The solid (dashed) black line shows the 90$\%$ (50$\%$) credible contour of the GRB localization. The gray points show the positions injected according to the GRB localization map to compute the sky-averaged value of $D_{90}^{\mathrm{TDR}}$. Both panels show results for GRB 200228291, the GRB with the highest value of $D_{90}^{\mathrm{TDR}}$ in our sample.}
    \label{fig:map}
\end{figure}

We derived the TDR for all the GRBs detected by the Fermi-GBM and Swift-BAT instruments during the first three observing runs O1-O2-O3a-O3b of LVK. The list of GRBs triggered by Fermi-GBM was taken from the FERMIGBRST catalog\footnote{\url{https://heasarc.gsfc.nasa.gov/w3browse/fermi/fermigbrst.html}}, and the list of GRBs triggered by Swift-BAT was taken from the online Swift/BAT Gamma-Ray Burst Catalog\footnote{\url{https://swift.gsfc.nasa.gov/results/batgrbcat/}}. When the burst was detected by both, we only considered the Swift catalog because it is better localized. The Fermi-GBM healpix map was retrieved using the \texttt{TriggerFinder} module from the Fermi GBM data tools\footnote{\url{https://fermi.gsfc.nasa.gov/ssc/data/analysis/gbm/gbm_data_tools/gdt-docs/index.html}}. Older Fermi-GBM healpix maps were downloaded from this repository\footnote{ \url{https://zenodo.org/records/6727152}}. No cut on the GRB duration was applied. Out of 574 GRBs found to intersect with the time windows of O1 to O3, 467 have at least one IFO taking data at the GRB time.

\subsection{Comparison with \texttt{PyGRB} exclusion distances}

The searches performed by LVK on GRB targets restrict the search to a time window, spatial position, and set of binary intrinsic parameters that are informed by the GRB detection itself. 
The modeled GRB targeted search is based on the use of the \texttt{PyGRB} pipeline. The targeted search exploits a reduced parameter space to perform a more sensitive search, making it possible to find possible GW candidates associated with the GRB, which would otherwise be missed by all-sky all-time searches. In the absence of a significant candidate, nondetection exclusion distances are derived. To first order, the exclusion distance is driven by the sensitivity of the interferometers and their angular response in the direction and at the time of the GRB. We therefore expect that the 90$\%$ exclusion distance found by \texttt{PyGRB} is fairly well approximated by the TDR derived here. 

We performed this comparison considering all the GRBs detected by Fermi and Swift and reported in the GRB targeted search papers for the O1 \citep{2017ApJ...841...89A}, O2 \citep{2019ApJ...886...75A}, O3a \citep{2021ApJ...915...86A}, and O3b \citep{2022ApJ...928..186A} observing runs, whose exclusion distance is estimated with \texttt{PyGRB}. 
According to the selection cuts adopted in the LVK analyses, the \texttt{PyGRB} search was performed on GRBs classified as short or ambiguous, namely for which there is a non-negligible chance of belonging to the short GRB class. 
For the comparison with \texttt{PyGRB}, we only considered GRBs for which the IFOs online we found matched those used in the LVK papers. For 71 out of 82 Fermi and Swift GRBs analyzed by \texttt{PyGRB}, we computed the TDR adopting the same IFOs. The mismatch occurs because either in the LVK analysis one IFO is added even though it is outside the nominal observing mode, or an IFO is excluded because it fails data-quality checks. For the distribution of the inclination angle, instead of using the default maximum value of 45 deg, as motivated in Sec.~\ref{em_prior}, we used 30 deg to be consistent with the injections used in the LVK works. While \texttt{PyGRB} exclusion distances are obtained using BNS injected with an NS mass distributed as a Gaussian with a mean of 1.4 $M_{\odot}$ and a standard deviation of 0.2 $M_{\odot}$, we obtained the TDR by fixing both masses at 1.4 $M_{\odot}$. This difference introduces only a negligible bias.

In contrast to the BNS case, the NSBH scenario is subject to substantially larger modeling uncertainties. LVK adopts a broad black hole mass distribution, whereas our framework assumes fixed component masses. Furthermore, different prescriptions (involving spins) are used to define the EM-bright region of the parameter space. These differences introduce systematic effects that are not straightforward to quantify or control in a direct comparison, and for this reason, we limited our analysis to BNS systems.

\subsection{Optimizing the match between the \texttt{PyGRB} exclusion distance and the TDR}
\label{optimize}

As mentioned above, the choice of the S/N threshold for the computation of $D_{90}^{\mathrm{TDR}}$ is arbitrary. We investigated whether any value of S/N threshold minimizes the difference between the TDR and the exclusion distance provided by \texttt{PyGRB}. For this purpose, we defined the quantity 
\begin{equation}
    \Delta_{90} (\rho_{\mathrm{cut}})=\left (D_{90}^{\mathrm{TDR}}(\rho_{\mathrm{cut}}) - D_{90}^{\mathrm{PyGRB}}\right )/D_{90}^{\mathrm{TDR}}(\rho_{\mathrm{cut}}),
\end{equation}
and we minimized the average of $|\Delta_{90}|$ over all the GRBs. The minimization was made using the optimal S/N and matched-filter S/N. Fig. \ref{fig:minim} shows the dependence of $|\Delta_{90}|$ on the S/N cut. The S/N cut was sampled with a precision of 0.5. $|\Delta_{90}|$ was minimized using a matched-filter (optimal) S/N cut of 9 (10).
The cumulative distribution of $\Delta_{90} (\rho_{\mathrm{cut}=9})$ is shown in Fig.~\ref{fig:cumul}. The median of the distribution is 0.02, and the 1$\sigma$ equivalent range is [-0.14,0.20]. Overall, this demonstrates that $D_{90}^{\mathrm{TDR}}(\rho_{\mathrm{cut}=9})$ is a good approximation of the \texttt{PyGRB} exclusion distance, even if with a non-negligible subset of cases where the mismatch between the two can reach values of 20 - 60 $\%$.

The residual difference can be ascribed to (i)  the non-stationary and non-Gaussian nature of real detector noise \citep{LIGOO4aDetChar}, (ii) differences between the injection populations used by \texttt{PyGRB} and our study, and (iii) the complexity of the template–injection mismatch, which exceeds what our simplified model can capture via the factor $k$. This demonstrates that even though the TDR tool is useful for a fast and cheap estimation of the GW detectability, the exclusion distance provided offline by targeted searches like \texttt{PyGRB} remains the most accurate and reliable metric to assess the possible connection between the EM transient and a CBC.

\subsection{TDR results}

In Fig.~\ref{fig:time_evo} we show the value of  $D_{90}^{\mathrm{TDR}}|_{\rho_{\mathrm{cut}}=9}$ for each GRB detected during one of the first three LVK observing runs, with a color-code indicating the geometrical network antenna factor, defined as
$F_{\mathrm{net}}=\sqrt{\sum_i\left(F_{+, i}^2+F_{\times, i}^2\right)},
$
where $F_{+}$,$F_{\times}$ are the plus and cross components of the antenna pattern of the interferometer, and the sum is extended to all IFOs online. This definition only includes the geometrical dependence on the external trigger position relative to each IFO and not the sensitivity of the IFO itself. The network antenna factor introduced here is therefore purly meant to demonstrate its correlation with the TDR, as discussed at the end of this section. When the GRB is localized by Swift-BAT, the antenna factor reported in the plot is computed at the sky position provided, while when it is localized by Fermi-GBM, the antenna factor is computed at the peak of the probability map. The horizontal dashed line indicates the median of $D_{90}^{\mathrm{TDR}}|_{\rho_{\mathrm{cut}}=9}$ over each observing run. The plot shows a clear improvement in sensitivity across the observing runs, rising from $\sim$70 Mpc in O1 to $\sim $140 Mpc at the end of O3, in agreement with the evolution of the sky-averaged BNS range  \citep{PhysRevD.93.112004,2021CQGra..38m5014D}. As expected, $D_{90}^{\mathrm{TDR}}$ is strongly correlated with the position of the GRB with respect to the antenna pattern, and also with the number of IFOs online. We report in Tab.~\ref{tab_grb} the list of all the GRBs analyzed here, with respective values of TDR for BNS and NSBH, including a comparison with \texttt{PyGRB} exclusion distances where available. As motivated in Sec.~\ref{meatch}, to be more on the conservative side, we report in the table the TDR adopting a $\rho_{\mathrm{cut}}=10$, while for the comparison with the \texttt{PyGRB} BNS exclusion distances, we use a $\rho_{\mathrm{cut}}=9$. 

Our results can be compared with \citep{2022ApJ...939L..14W}, who performed a search, using PyCBC, of GW candidates in temporal proximity to long GRB detected during O3. They also reported nondetections, along with exclusion distances, defined as the distance where a BNS with 1.4+1.4 $M_{\odot}$ would have an optimal S/N = 8, fixing the inclination angle at 30 deg. In order to perform a consistency check, we ran the TDR on the same sample of GRBs and assumed the same injection parameters. The results are consistent up to a few percent. 
It is informative to also compare our results with \citealt{2026MNRAS.549ag834S}, who also performed a search of GW candidates in coincidence with long GRBs detected during O3. The authors assessed the presence of a CBC candidate by performing a full parameter estimation and computing the correspondent Bayesian evidence. No significant coincidence was found, and 90$\%$ exclusion distances were derived for all the GRBs. Since the authors followed a Bayesian approach, the value of the 90$\%$ exclusion distance depends on the choice of the distance prior, which was uniform in volume in their case. Their study assumed a component mass prior that encompassed both BNS and low-mass NSBH. Nonetheless, their 90$\%$ exclusion distance is systematically larger on average than our BNS and NSBH $D_{90}^{\rm TDR}$. The distance prior weights more higher values of the distance, which explains why an exclusion distance obtained with a Bayesian approach is systematically larger than the distance derived here. 

\subsection{TDR standard products}

As an example, we show the standard products of the TDR tool for the GRB with the highest value of $D_{90}^{\mathrm{TDR}}$ in our sample, corresponding to GRB200228291. The first product, shown in Fig.~\ref{fig:map} (top panel), corresponds to the fraction of injected BNS sources recovered with a $\rho_{\mathrm{MF}}>9$ as a function of luminosity distance. The result is reported for different values of component masses and two prior assumptions on the inclination angle. The central panel reports the same for a NSBH system. The second product, shown in Fig.~\ref{fig:map} (bottom panel), reports the value $D_{90}^{\mathrm{TDR}}$ as a function of the sky position. The localization contours of the GRB are shown as well, together with the injected positions distributed according to the probability map itself. The spatial dependence of the $D_{90}^{\mathrm{TDR}}$ directly correlates with the network antenna pattern. The high value of $D_{90}^{\mathrm{TDR}}$ is indeed motivated by the GRB localization being mostly concentrated around one of the peaks of the antenna pattern response of the IFOs.

\section{Examples of science cases for TDR}
\label{cases}
The TDR use cases address two complementary objectives: enabling the low-latency optimization of follow-up strategies and the prioritized allocation of observational resources, and interpreting the observations. In the latter case, by combining GW observations with complementary information, such as the temporal and spectral source properties and the characteristics of its host galaxy, the TDR use cases help us to assess the nature of the source, and in particular, support or rule out a CBC origin. 
We summarize some possible science cases where the TDR can be beneficial below.

1) Inform and optimize EM follow-up observations. Two illustrative examples are GRB 250225B \citep{2025GCN.39473....1W} and GRB 251026B \citep{2025GCN.42455....1F}. For GRB 250225B, detected by \textit{Swift}-BAT, a galaxy at 78 Mpc was identified within the arcminute localization region, with a chance coincidence probability of $<$$1\%$ \citep{2025GCN.39494....1S,2025GCN.39495....1Y}. A prompt estimate of the GW detector sensitivity at the distance of the candidate host would have strengthened or weakened the merger interpretation, and in the absence of a GW detection despite a high detection probability, would have strongly disfavored the proposed host galaxy. Similarly, for the short GRB 251026B, detected by \textit{Fermi}-GBM, GOTO identified a candidate optical counterpart in a galaxy at $\sim$250 Mpc \citep{2025GCN.42470....1B}, later classified as a Type Ia supernova \citep{2025GCN.42519....1C}. In this case, the prompt dissemination of the TDR GW exclusion distance would have rapidly disfavored the merger scenario. More generally, comparing the host-galaxy distance with the GW exclusion distance provides an efficient criterion for prioritizing EM follow-up: host galaxies within the GW exclusion distance make a merger origin unlikely, whereas more distant hosts leave the merger scenario viable and justify continued observational effort when the goal is to identify merger-driven events.

2) Interpret the origin of EM transients. The recent observations of GRB 211211A \citep{2022Natur.612..223R,2022Natur.612..236M,2022Natur.612..232Y,2024Natur.626..742Y} and GRB 230307A \citep{2023arXiv230800633G,2024Natur.626..742Y,2024Natur.626..737L,2025ApJ...994...17D,2025NSRev..12E.401S} have raised the possibility that some long-duration GRBs, typically associated with massive star collapse, might instead originate from CBCs \citep{2026arXiv260726133L}. Moreover, the distance of their host galaxies, $\sim$350 Mpc \citep{2022Natur.612..223R} and $\sim$290 Mpc \citep{2024Natur.626..737L}, respectively, places these two GRBs within the current sensitivity horizon of the LVK network, especially when powered by NSBH systems \citep{2023ApJ...958L..33G}. In this context, the TDR can help us to identify potential merger-driven events among long-duration GRBs by providing a rapid assessment of their GW detectability. When the absence of significant GW search pipeline candidates is combined with complementary astronomical observations, the assessment of GW detectability remains important beyond the early stages of EM follow-up. Furthermore, the TDR can provide constraints on the physical interpretation of poorly understood transients, including the emerging class of fast X-ray transients \citep{Chopra:2026skb}, for which Einstein Probe is playing a pivotal role through their systematic discovery \citep{2019Natur.568..198X,2019ApJ...886..129S,2024A&A...683A.243Q,2024A&A...690A.101W,2026MNRAS.545f2021J,2026arXiv260627048V}.

3) Derive constraints on the CBC origin of the EM source when the merger time is not well known. This is particularly relevant for optical transients whose early-time evolution resembles that of a kilonova, as in the case of the ZTF candidate AT2025ulz \citep{2025GCN.41414....1S,2025ApJ...995L..27G,2025ApJ...995L..59K,2025ApJ...995L..47O,2025arXiv251024620H,2026ApJ..1001L..20H,2026arXiv260502639A}, found in coincidence with the low-significance GW candidate S250818k \citep{2025GCN.41437....1L}. The transient was later classified as a Type II supernova \citep{2026arXiv260502639A} and was also proposed as a candidate for the superkilonova scenario \citep{2007ApJ...658.1173P,2024ApJ...971L..34M,2025ApJ...995L..59K}. Although AT2025ulz was discovered in response to a search targeted on a GW candidate, similar optical transients might be serendipitously identified in wide-field surveys without any accompanying GW trigger. By modeling the early-time light curve under the hypothesis of a merger-driven optical transient, an estimate of the putative merger time can be inferred. Even in the presence of substantial temporal uncertainty, the TDR framework enables a rapid and systematic exploration of the GW detection probability across the whole time window. For sources similar to AT2025ulz, the TDR can be useful to provide informative constraints on the controversial nature of the transient. Analogous considerations apply to orphan afterglow \citep{2015A&A...578A..71G}. In these cases, multiwavelength observations reveal features consistent with a GRB jet, even though the high-energy prompt emission is not detected. This nondetection can arise from observational limitations, an intrinsically subluminous prompt emission, or a jet that is misaligned with respect to the line of sight. By modeling the multiwavelength light curve of the candidate, it is possible to estimate the jet launch time \citep{2020ApJ...905...98H,2021ApJ...918...63A,2022ApJ...938...85H,2026ApJ..1000..118G}. In the case of well-sampled afterglow light curves, the associated uncertainty can be as small as $\sim$ 200-300 s \citep{2023A&A...679A.103X,2025MNRAS.538..351S,2025MNRAS.537.2362P,2025ApJ...985..124L}. The TDR analysis can then be performed over the full range of this uncertainty, disfavoring or leaving open the merger origin channel.

\section{Conclusions}
\label{conc}
We have demonstrated how the observational properties of an astronomical transient can be exploited to compute the GW detectability range under the assumption of a compact binary merger origin. Along with a description of the method, we provided an illustrative application of this tool to all the short- and long-duration GRBs detected during the first three observing runs of LVK. By defining the targeted detectability range (TDR) as the distance where 90$\%$ of the injected sources have a matched-filter S/N $>$ 9, the derived values are close to the 90$\%$ exclusion distance derived by \texttt{PyGRB}, with a relative mismatch that is $\lesssim20\%$ for $\sim70\%$ of the analyzed GRBs. In the absence of GW candidates found by low-latency all-sky searches in temporal proximity to the external trigger, the TDR can be interpreted as an exclusion distance informed by the observational features of the astrophysical source.

The TDR tool can run with minimal computational effort and deliver a range estimate within minutes of the EM trigger. In addition to GRBs, this tool can be systematically applied to kilonova candidates, fast X-ray transients, fast radio bursts, fast blue optical transients, and high-energy neutrinos when an estimate of the trigger time and sky localization is provided by the EM facility. A public low-latency release of TDR results would be of broad benefit to the astronomical community, providing direct access to robust estimates of the GW detection probability associated with any given external messenger. This is fundamental for prioritizing  EM follow-up campaigns and efficiently allocating observational resources. 

Even in the presence of significant uncertainty regarding the delay between the merger time and the EM trigger, the TDR has the substantial advantage that it can be estimated sequentially on multiple time segments before and after the external trigger, probing the time evolution of the sensitivity of the GW interferometers and deriving the most conservative and robust conclusions about the merger origin of the transient. This feature is crucial for the systematic application of TDR on all the fast X-ray transients detected during the second half of the fourth LVK observing run, as detailed in a companion paper \citep{Chopra:2026skb}.

Looking ahead, as LVK detectors will enter the fifth observing run (O5), the GW horizon will expand through the improved sensitivity, enabling the use of the TDR tool to place increasingly stringent constraints on the growing sample of EM transients expected to be discovered by wide-field surveys such as the Vera Rubin Observatory and Einstein Probe. This gain will further be amplified in the era of next-generation detectors such as the Einstein Telescope \citep{2010CQGra..27s4002P} and Cosmic Explorer \citep{2021arXiv210909882E}, whose reach will extend the CBC detection to higher redshifts.
Their unprecedented sensitivity will allow us not only to obtain a systematic detection of CBCs, but also to exploit the nondetection of GWs in coincidence with EM triggers to place constraints on the physical conditions that favor the EM emission.

\section*{Data Availability}
The full version of Table \ref{tab_grb} is available at the following Zenodo link: \url{https://doi.org/10.5281/zenodo.20274743}.
\begin{acknowledgements}
      M.B. and S.R. acknowledge support from the Astrophysics Center for Multi-messenger Studies in Europe (ACME), funded under the European Union’s Horizon Europe Research and Innovation Program, Grant Agreement No. 101131928. The authors thank Matteo Pracchia, Manuel Arca Sedda and Filippo Santoliquido for the useful comments.
\end{acknowledgements}

\begin{table*}[]
  \centering
  \caption{Relevant parameters used for the waveforms in our analysis.}
  \label{tab:spin_thresholds}
  \begin{tabular}{ccccccccc}
    \toprule
    $M_{\mathrm{NS}}$ ($M_\odot$) & $M_{\mathrm{BH}}$ ($M_\odot$) &
    $q$ & $C_{\mathrm{SFHo}}$ & $\Lambda_{\mathrm{SFHo}}$ &  $\chi_{\mathrm{BH,min}}^{\mathrm{SFHo}}$ &
    $C_{\mathrm{DD2}}$ & $\Lambda_{\mathrm{DD2}}$ & $\chi_{\mathrm{BH,min}}^{\mathrm{DD2}}$ \\
    \midrule
    1.00 & 5.00 & 5.00 & 0.1226 & 3160 & 0.00000 & 0.1126 & 4970 & 0.00000 \\
    1.40 & 10.00 & 7.14 & 0.1734 & 436 & 0.64638 & 0.1559 & 829 & 0.51135 \\
    2.00 & 20.00 & 10.00 & 0.2683 & 16 & 0.98617 & 0.2238 & 76 & 0.93732 \\
    \bottomrule
  \end{tabular}
\tablefoot{For each combination the compactness $C$ is calculated from the respective EOS, the minimum black hole spin required to obtain non-zero ejecta mass following \citet{Foucart2018} and the tidal deformability $\Lambda$ is derived using formulae from \citet{Godzieba2021}. The parameter $q$ indicates the mass ratio.}
\end{table*}

\begin{table*}

\centering
\caption{Observation runs with detector channels used for each interferometer.}
\begin{tabular}{l c c c c c}
\toprule
Run & Start (GPS) & End (GPS) & H1 & L1 & V1 \\
\midrule

O1  & 1126051217 & 1137254417 &
\makecell[l]{H1:DCS-CALIB\_\\STRAIN\_C02} &
\makecell[l]{L1:DCS-CALIB\_\\STRAIN\_C02} &
-- \\
\midrule

O2  & 1164556817 & 1187733618 &
\makecell[l]{H1:DCH-CLEAN\_\\STRAIN\_C02} &
\makecell[l]{L1:DCH-CLEAN\_\\STRAIN\_C02} &
\makecell[l]{V1:Hrec\_hoft\_\\V1O2Repro2A\_16384Hz} \\
\midrule

O3a & 1238166018 & 1253977218 &
\makecell[l]{H1:DCS-CALIB\_\\STRAIN\_CLEAN\_\\SUB60HZ\_C01} &
\makecell[l]{L1:DCS-CALIB\_\\STRAIN\_CLEAN\_\\SUB60HZ\_C01} &
\makecell[l]{V1:Hrec\_hoft\_16384Hz} \\
\midrule

O3b & 1256655618 & 1269363618 &
\makecell[l]{H1:DCS-CALIB\_\\STRAIN\_CLEAN\_\\SUB60HZ\_C01} &
\makecell[l]{L1:DCS-CALIB\_\\STRAIN\_CLEAN\_\\SUB60HZ\_C01} &
\makecell[l]{V1:Hrec\_hoft\_16384Hz} \\

\bottomrule
\end{tabular}
\tablefoot{A detailed description of frames nomenclature and properties is available in \cite{2026arXiv260527223T}.}

\label{runs}
\end{table*}

\begin{table*}[]
    \centering
    \caption{Summary of TDR ($D_{90}^{\mathrm{TDR}}$) values for all the GRBs detected during the first three LVK observing runs.}
    \begin{tabular}{llclrrrr}
\toprule
GRB & Instrument & Trigger time & IFO 
& \multicolumn{3}{c}{BNS} 
& NSBH \\
\cmidrule(lr){5-7}
& & & & 
$D_{90}^{\mathrm{TDR}}|_{\rho_{\mathrm{cut}}=10}$ 
& $D_{90}^{\mathrm{TDR}}|_{\rho_{\mathrm{cut}}=9}$ 
& $D_{90}^{\rm pyGRB}$ 
& $D_{90}^{\mathrm{TDR}}|_{\rho_{\mathrm{cut}}=10}$ \\
\midrule
& & (UTC)& & (Mpc) & (Mpc) & (Mpc) & (Mpc) \\
\midrule
GRB150913161&Fermi&2015-09-13T03:51:57&H1&41&47&-&91\\
GRB150922234&Fermi&2015-09-22T05:37:29&H1L1&79&90&-&162\\
GRB150922883&Fermi&2015-09-22T21:11:32&L1 (H1L1)&37&42&71&83\\
GRB150923297&Fermi&2015-09-23T07:07:36&H1L1&89&101&98&181\\
GRB150923429&Fermi&2015-09-23T10:18:17&H1L1&117&133&136&242\\
\bottomrule

    \end{tabular}
    \tablefoot{The TDR is reported by default using a S/N cut at 10. The comparison with \texttt{PyGRB} exclusion distances is done for BNS systems and adopting an S/N cut of 9. $D_{90}^{\mathrm{TDR}}$ values  for BNS are computed assuming $m_1=m_2=1.4\,M_\odot$ and for NSBH $m_{\mathrm{BH}}=10\,M_\odot$, $m_\mathrm{NS}=1.4\,M_\odot$. The $D_{90}^{\mathrm{TDR}}$ values for NSBH show the average obtained assuming SFHo and DD2 EOSs. The inclination angle of the binary is assumed to be isotropically distributed between 0 and 30 deg. A dash indicates that no \texttt{PyGRB} exclusion distance is available, since the GRB is classified as long. If the set of IFOs used in our analysis differs from that used in LVK targeted searches, the latter is reported in brackets. GRB170817529 corresponds to GRB170817A, detected in coincidence with GW170817. Full table available online at the link reported in the data availability section.}
    \label{tab_grb}
\end{table*}

\bibliographystyle{aa}
\bibliography{ref}

\end{document}